\documentclass[aps,prd,preprintnumbers,nofootinbib,onecolumn,superscriptaddress]{revtex4-2}
\usepackage[colorlinks,linkcolor=blue,anchorcolor=violet,citecolor=red]{hyperref}
\usepackage{amsmath,amssymb}
\usepackage[dvipdfmx]{graphicx}
\usepackage{float}
\usepackage{comment}
\usepackage{xcolor}
\usepackage{bm}

\usepackage{ascmac}
\usepackage{physics}
\usepackage{color,float}
\usepackage{enumerate}

\begin{document}

\title{Consistency between causality and complementarity guaranteed by Robertson inequality in quantum field theory 
}%

\author{Yuuki Sugiyama}
 \email{sugiyama.yuki@phys.kyushu-u.ac.jp}
\affiliation{Department of Physics, Kyushu University, 744 Motooka, Nishi-Ku, Fukuoka 819-0395, Japan}

\author{Akira Matsumura}
 \email{matsumura.akira@phys.kyushu-u.ac.jp}
\affiliation{Department of Physics, Kyushu University, 744 Motooka, Nishi-Ku, Fukuoka 819-0395, Japan}
 
\author{Kazuhiro Yamamoto}
 \email{yamamoto@phys.kyushu-u.ac.jp}
\affiliation{Department of Physics, Kyushu University, 744 Motooka, Nishi-Ku, Fukuoka 819-0395, Japan}
\affiliation{
Research Center for Advanced Particle Physics, Kyushu University, 744 Motooka, Nishi-ku, Fukuoka 819-0395, Japan}

\begin{abstract}
It has long been debated whether gravity should be quantized or not.
Recently, the authors in \cite{Mari, Baym} discussed the inconsistency between causality and complementarity in a Gedankenexperiment involving the quantum superposition of massive/charged bodies, 
and Belenchia et al. \cite{Belenchia2018, Belenchia2019} 
resolved the inconsistency by requiring the quantum radiation and vacuum fluctuations of gravitational/electromagnetic field.
Stimulated by their works, we reanalyze the consistency between the two physical properties, causality and complementarity, according to the quantum field theory. 
In this analysis, we consider a Gedankenexperiment inspired by  \cite{Mari,Baym, Belenchia2018,Belenchia2019}, in which two charged particles coupled with a photon field are in a superposition of two trajectories. 
First, we observe that causality is satisfied by the retarded propagation of the photon field. 
Next, by introducing an inequality between visibility and which-path information, we show that the quantum radiation and vacuum fluctuations of the photon field ensure complementarity. 
We further find that the Robertson inequality associated with the photon field leads to the consistency between causality and complementarity in our Gedankenexperiment.
Finally, we mention that a similar feature appears in the quantum field of gravity.
\end{abstract}

\maketitle
\section{Introduction\label{intro}}
The unification of quantum mechanics and general relativity is a fundamental unsolved problem in theoretical physics.
Despite all the efforts that have been made, the exact theory of quantum gravity has not yet been completed.
Moreover, we do not even know whether gravity really follows the principle of quantum mechanics or not \cite{KYFeynman,KYPenrose,Diosi}.
Recently, testing the quantum nature of gravity has attracted significant interest in theoretical physics, stimulated by the proposal by Bose et al. \cite{Bose2017}, and Marlleto and Vedral \cite{Marlleto2017}. 
The BMV proposal suggests that quantum entanglement due to the Newtonian potential between two masses can be an evidence of quantum gravity, which can be tested by a tabletop experiment (see also \cite{KYCarney2019}).  
Inspired by their works \cite{Bose2017, Marlleto2017}, Newtonian entanglement was evaluated in experimental proposals for matter-wave interferometry \cite{Nguyen2020, Miki2021},
mechanical oscillators \cite{Krisnanda2020, Qvafort2020}, optomechanical systems \cite{Balushi2018,Miao2020,Matsumura2020,Miki2022},
hybrid systems \cite{Carney2021a,Pedernales2021, LG, Streltsov}, etc.
However, there is room for arguments to understand what the detection of the Newtonian entanglement means, e.g., how the Newtonian entanglement is
related to the quantum field theory of gravity and gravitons
\cite{Hu,KYMarshman,KYAnastopoulos,KYCarney2,KYBose2,Danielson2021}.

We revisit the entanglement generation in the BMV proposal in the framework of the quantum field theory by focusing on a paradox in a Gedankenexperiment, which was previously analyzed in
Refs.~\cite{Mari,Baym,Belenchia2018, Belenchia2019,Danielson2021}.
In the Gedankenexperiment (see Fig.~\ref{fig:setup}), Alice prepares a particle in a superposition of spatially localized states separated by a distance $L$ and starts to recombine her particle at a time 
$t=t_0$ to observe its interference. 
The recombination process is performed during a time 
$T_\text{A}$.
Bob, who is at a distance $D$ ($\gg L$) from Alice, can choose whether to release a particle at the time $t=t_0$. 
When Bob released his particle, after a time 
$T_\text{B}$, he measures his particle to determine the strength of the Newtonian/Coulomb force induced by Alice's particle and gains information about which path her particle took.
The actions of Alice and Bob after the time $t=t_0$ occur in spacelike separated regions ($D>T_\text{A}$ and $D>T_\text{B}$).
If Bob acquires any which-path information from his measurement, the state of his particle must
be entangled with Alice's particle. 
This leads to the correlation between Alice and Bob.
Then, because of the correlation due to the entanglement,  Alice's particle cannot be in a perfect coherent superposition when Bob measures his released particle.
This is the result of complementarity. 
However, when Bob does not release his particle, Alice's particle can maintain perfect coherence. Bob's choice affects the coherence of Alice's particle. 
Since Alice and Bob perform their actions in a spacelike separated region, it is
impossible for Bob's measurement to have any effect on Alice's result owing to causality.
This leads to the apparent violation of causality or complementarity.
This paradox was first discussed in Ref.~\cite{Mari,Baym}, and the authors in Refs.~\cite{Belenchia2018, Belenchia2019,Danielson2021} claimed that the paradox
can be resolved by Alice's limitation in maintaining coherence due to the emission of entangling gravitons/photons during the process of recombination of her particle and Bob's limitation
in acquiring which-path information due to the vacuum fluctuations of gravitational/electromagnetic field (for a brief review, see Sec. \ref{gedanken}).
The most important implication made by the above mentioned authors is that the existence of a quantum gravitational field and gravitons may be necessary to solve the paradox.

In this study, we reanalyze the paradox rigorously by estimating the feasibility of the measurements by Alice and Bob. 
We use the theoretical model developed in \cite{Sugiyama}, in which we investigated entanglement generation between a pair of charged particles in a superposition of spatially localized states based on quantum electrodynamics.
We demonstrate that the causality in our model is automatically satisfied by the retarded propagation of the photon field. 
Furthermore, by estimating the visibility measured by Alice and the distinguishability in Bob's
measurement, we show that the complementarity in our model is protected by the radiation and vacuum fluctuations of the photon field. 
Additionally, we prove that the complementarity is guaranteed by the Robertson inequality for the photon field, which reflects the non-commutativity of a quantized field. 
From the analogy between electromagnetic dynamics and general relativity, we mention that a similar feature may appear in quantum gravitational fields.

The remainder of this paper is organized as follows. 
In Sec. \ref{gedanken}, we briefly review the paradox in the Gedankenexperiment by following Refs.~\cite{Belenchia2018, Belenchia2019}.
In Sec. \ref{causality}, we demonstrate that causality is not violated. 
In Sec. \ref{complementarity}, we show that complementarity is satisfied for two charged particles coupled with a photon field.
Section \ref{conclusion} is devoted to the summary and conclusion.
In Appendix \ref{trace}, we derive Eqs.~\eqref{rhoA} and \eqref{rhoB}.
In Appendix \ref{proof}, we prove the inequality in visibility and distinguishability.
In Appendix \ref{proofin}, we present the proof of the statement in \eqref{suffcon}.
Throughout this study, we used the natural units with $c=\hbar=1$.

\section{A brief review of the Gedankenexperiment\label{gedanken}}
\begin{figure}[H]
  \centering
  \includegraphics[width=0.6\linewidth]{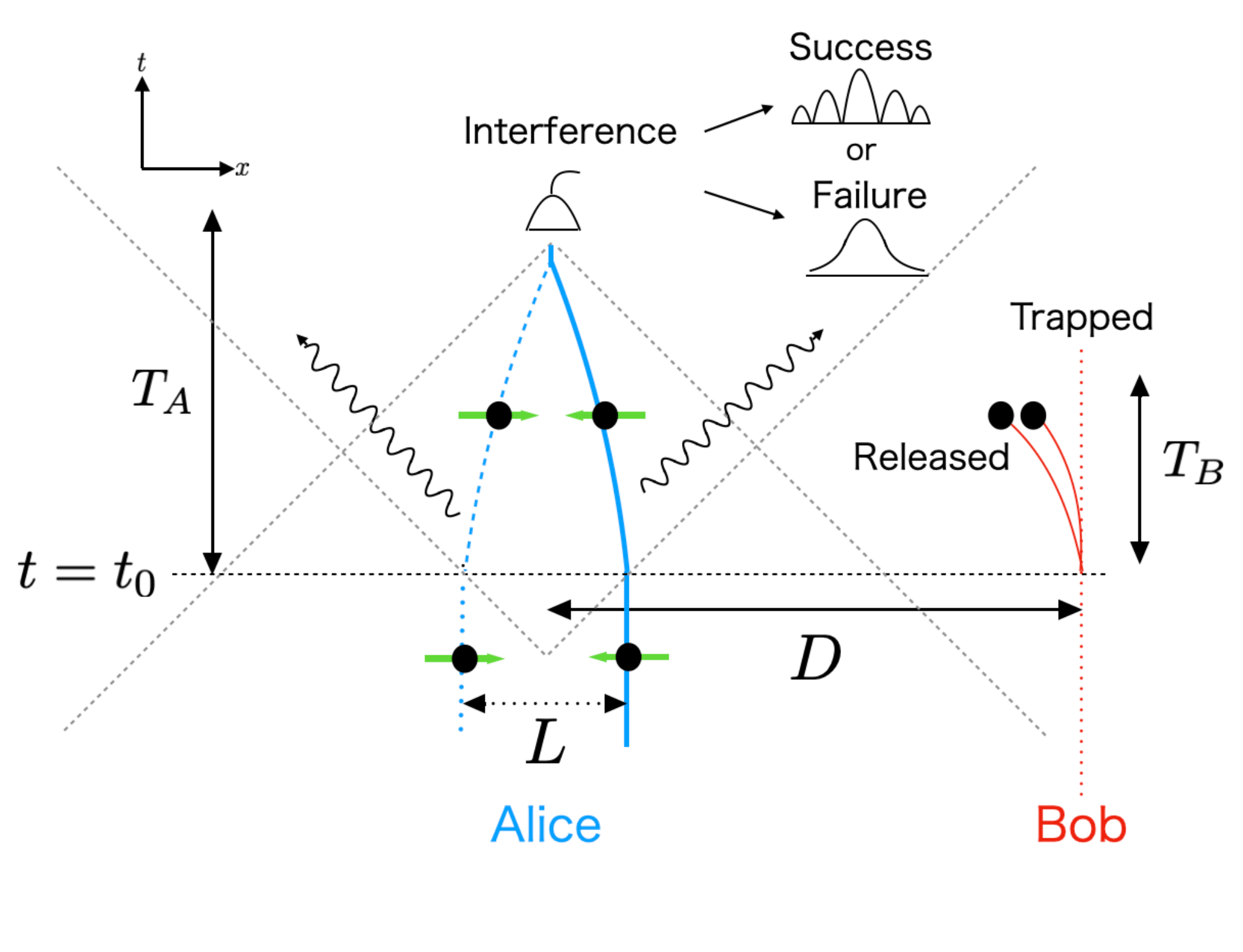}
  \caption{Setup for the Gedankenexperiment introduced by \cite{Belenchia2018, Belenchia2019}.}
  \label{fig:setup}
\end{figure}

In this section, we review the paradox of the Gedankenexperiment addressed in Refs.~\cite{Mari,Baym, Belenchia2018, Belenchia2019,Danielson2021}.
As is shown in Fig.~\ref{fig:setup}, Alice and Bob are separated by a distance $D$. 
Their particles interact via the Newtonian/Coulomb potential.
Alice's particle with a spin is in a superposition of spatially localized states
separated by a distance $L$, which was prepared through a Stern-Gerlach apparatus,
and an interference experiment is performed during a time $T_{\text{A}}$.
In contrast, Bob chooses whether his particle is released or trapped at a time $t=t_0$.
If Bob releases his particle, it moves under the gravitational/electromagnetic potential created by Alice's particle. 
After a time $T_\text{B}$, he measures the position of his particle.
Assuming the regimes $D > T_{\text{A}}$ and $D> T_{\text{B}}$, in which Alice and Bob perform their actions in spacelike separated regions, we can consider the following two incompatible arguments.
\begin{enumerate}[(i)]
\item
If causality holds, Alice can observe the interference pattern of her particle regardless of whether Bob measures his particle.
\item 
If complementarity holds, Bob's measurement of his particle should lead to the decoherence of Alice's particle.
\end{enumerate}
Arguments (i) and (ii) seem to contradict each other, and thus the paradox appears.

The authors in Refs. \cite{Belenchia2018, Belenchia2019,Danielson2021} claimed that this paradox is resolved by the quantum radiation of gravitons/photons emitted by massive/charged particles and the vacuum fluctuations of gravitational/electromagnetic fields.
The quantum radiation from Alice's particle causes the decoherence of her particle, and then the interference experiment fails.
In other words, this entangling radiation limits the maintenance of coherence in Alice's experiment.
The presence of the vacuum fluctuations limits the ability to obtain the which-path information of Alice's particle for Bob's measurement.
The two effects, the decoherence due to quantum radiation and the limitation of which-path information due to vacuum fluctuations are key to resolving this paradox  \cite{Belenchia2018, Belenchia2019, Danielson2021}.

In the following two sections, we reanalyze the consistency between causality and complementarity by 
assuming a situation similar to that in Fig. \ref{fig:setup}. 
This is an extension of a previous study
\cite{Sugiyama}, which investigated the effect of vacuum fluctuations of a photon field on the electromagnetic version of the BMV proposal. 
This work is based on the quantum electromagnetic dynamics; however, our result can be
reinterpreted for the quantized gravitational field, as discussed in Sec. V.

\section{consistency of causality\label{causality}}
In this section, we show that Bob's particle does not affect Alice's particle because of the causality satisfied for $D> T_A$ and $D>T_B$. 
We first introduce the model of two charged particles (Alice's particle and Bob's particle) coupled with a photon field developed in Ref.~\cite{Sugiyama}.
The total Hamiltonian of our system is composed of the local Hamiltonians of each charged particle 
$\hat{H}_{\text{A}}$ and $\hat{H}_{\text{B}}$, the free Hamiltonian of the photon field 
$\hat{H}_\text{ph}$, and the interaction term $\hat{V}$ as
\begin{eqnarray}
\hat{H}=\hat{H}_{\text{A}}+\hat{H}_{\text{B}}+\hat{H}_\text{ph}+\hat{V}, 
\quad 
\hat{V}=\int d^3x \Big(\hat{J}^\mu_{\text{A}}(\bm x)+\hat{J}^\mu_{\text{B}}(\bm{x})\Big)\hat{A}^{\mu}(\bm{x}),
\end{eqnarray}
where 
$\hat{J}^\mu_{\text{A}} $ 
and 
$\hat{J}^\mu_{\text{B}} $ are the current operators of each particle coupled with the photon field operator $\hat{A}^\mu $.
\begin{figure}[htbp]
  \includegraphics[width=0.6\linewidth]{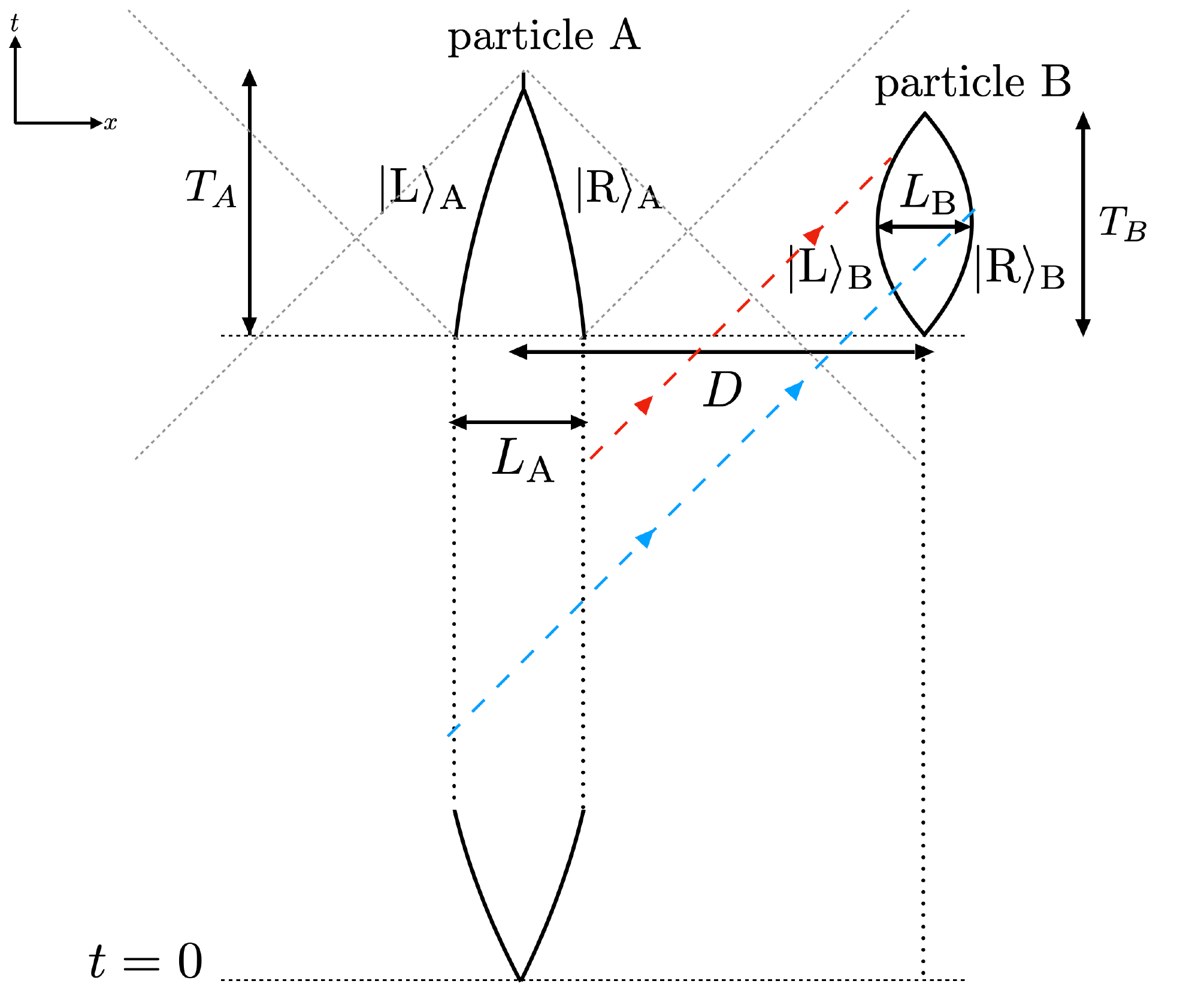}
\caption
{
  Configuration of our model.
  $L_{\text{A}}$ and $L_{\text{B}}$ are each separation of a spatial superposition of particles A and B, 
  and $D$ is a distance between Alice's system and Bob's system. 
  $T_{\text{A}}$ is a time scale recombining particle A,
  and particle B in Bob's system is superposed during a time $T_{\text{B}}$.
  Here, we assume the regimes $D  > T_\text{A}$ and $D>T_\text{B}$.
 Particle A takes the right or left path $|\text{R}\rangle_{\text{A}}$($|\text{L}\rangle_{\text{A}}$) and induces the retarded photon field along each path (as shown in the dashed red or blue line). 
 The retarded field affects particle B moving the left or right path $|\text{L}\rangle_{\text{B}}$ and $|\text{R}\rangle_{\text{B}}$ .\\
  \label{fig:configuration}
}
\end{figure}
We consider the following initial condition
\begin{align}
|\Psi(0)\rangle
&=
\frac{1}{{2}}|\text{C}\rangle_\text{A}
(\ket{\uparrow}_{\text{A}}+\ket{\downarrow}_{\text{A}})
|\text{C}\rangle_\text{B}
(\ket{\uparrow}_{\text{B}}+\ket{\downarrow}_{\text{B}})
|\alpha\rangle_{\text{ph}},
\label{inistate2}
\end{align}
where $\ket{\uparrow}_{j}(\ket{\downarrow}_{j})$ are the spin degrees of freedom of the charged particle $j$ with $j=\text{A}, \text{B}$, and $|\text{C}\rangle_\text{A}$ and $|\text{C}\rangle_\text{B}$ denote
the localized particle wave function of A and B, respectively.
The photon field is in a coherent state $|\alpha\rangle_{\text{ph}}$ with $|\alpha \rangle_\text{ph}=\hat{D}(\alpha)|0\rangle_\text{ph}$.
$|0\rangle_\text{ph}$ is the vacuum state satisfying
$\hat{a}_\mu (\bm{k})|0 \rangle_\text{ph}=0$ for annihilation operator of the photon field $\hat{a}_\mu (\bm{k})$, and 
$\hat{D}(\alpha)$ is the unitary operator called a displacement operator defined as
\begin{equation}
\hat{D}(\alpha)=\exp \left[
\int d^3k (\alpha^\mu (\bm{k})\hat{a}^\dagger_\mu (\bm{k})-h.c.)
\right],
\label{D}
\end{equation}
where the complex function 
$\alpha^\mu (\bm{k})$ characterizes the amplitude and phase of initial photon field. 
The form of the complex function 
$\alpha^\mu (\bm{k})$ is restricted by the auxiliary condition in the BRST formalism~\cite{Sugiyama}.
The coherent state $|\alpha\rangle_{\text{ph}}$ is interpreted as a state in which there is a mode of the electromagnetic field following Gauss's law due to the presence of charged particles (See Appendix A of Ref.~\cite{Sugiyama}).
For $t<0$, the charged particles A and B are localized around each trajectory, whose states are described by $|\text{C}\rangle_\text{A}$ and $|\text{C}\rangle_\text{B}$, respectively. 
Then the photon field for $t<0$ is not in a quantum superposition and behaves classically.
In this case the states of $\text{A}$ and $\text{B}$ are uncorrelated with the photon field. Now, we assume that each particle
is manipulated through an inhomogeneous magnetic field ($|\text{C} \rangle_j \ket{\uparrow}_j \rightarrow |\psi_\text{L} \rangle_j \ket{\uparrow}_j, |\text{C} \rangle_j \ket{\downarrow}_j \rightarrow |\psi_\text{R} \rangle_j \ket{\downarrow }_j$) to create spatially  superposed states with $|\psi_\text{L}\rangle_{j}\ket{\uparrow}_{j}$, and $|\psi_\text{R}\rangle_{j}\ket{\downarrow}_{j}$, which is understood as the Stern–Gerlach effect discussed in~\cite{Belenchia2018, Bose2017}. 
In our Gedankenexperiment shown in Fig.\ref{fig:configuration}, each particle is spatially superposed at different times.
In the following, $|\text{C} \rangle_{j} \ket{\uparrow}_{j}$ and $|\text{C}\rangle_{j} \ket{\downarrow}_{j}$ are represented by $|\text{L}\rangle_{j}$ and $|\text{R}\rangle_{j}$ with $j=\text{A}, \text{B}$ for simplicity. 
The initial state is rewritten as 
\begin{align}
|\Psi(0)\rangle
&=
\frac{1}{{2}}
(|\text{L}\rangle_{\text{A}}+|\text{R}\rangle_{\text{A}})
(|\text{L}\rangle_{\text{B}}+|\text{R}\rangle_{\text{B}})
|\alpha\rangle_{\text{ph}},
\label{inistate22}
\end{align}
We note that $|\text{R}\rangle_{\text{A}}$ ($|\text{R}\rangle_{\text{B}}$) and $|\text{L}\rangle_{\text{A}}$ ($|\text{L}\rangle_{\text{B}}$) are the states of wave packets localized around classical trajectories. 
After each particle has passed through an inhomogeneous magnetic field, the states $|\text{L}\rangle_{j}$ and $|\text{R}\rangle_{j}$ are regarded as the localized states of the particle $j=\text{A},\text{B}$ around the left trajectory and the right trajectory shown in Fig.\ref{fig:configuration}, respectively.
We assume that the current operators $\hat{J}^\mu_{i\text{I}}(x)=e^{i\hat{H}_0 t} \hat{J}^\mu_{i}(0,\bm{x})e^{-i\hat{H}_0 t}$ in the interaction picture with respect to 
$\hat{H}_0 =\hat{H}_{\text{A}} + \hat{H}_{\text{B}} +\hat{H}_\text{ph}$ are approximated using the classical currents as
\begin{align}
&\hat{J}_{\text{AI}}^ \mu(x)\left|{\text{P}}\right\rangle_{\text{A}}
\approx J_{\text{AP}}^{\mu}(x)\left|{\text{P}}\right\rangle_{\text{A}},
\quad 
\hat{J}_{\text{BI}}^\mu(x)\left|{\text{Q}}\right\rangle_{\text{B}} \approx J_{\text{BQ}}^{\mu}(x)\left|{\text{Q}}\right\rangle_{\text{B}},
\label{approx3}
\\
&J_{\text{AP}}^{\mu}(x)
=e_\text{A} \int d \tau \frac{d X_{\text{AP}}^{\mu}}{d \tau} \delta^{(4)}\left(x-X_{\text{AP}}(\tau)\right),
\quad 
J_{\text{BQ}}^{\mu}(x)
=e_\text{B} \int d \tau \frac{d X_{\text{BQ}}^{\mu}}{d \tau} \delta^{(4)}\left(x-X_{\text{BQ}}(\tau)\right),
\label{approx4}
\end{align}
where 
$X^\mu_{\text{AP}}(\tau)$ and 
$X^\mu_{\text{BQ}} (\tau)$ with $\text{P}, \text{Q}=\text{R}, \text{L}$ represent the trajectories of each particle with coupling constants $e_{\text{A}}$ and $e_{\text{B}}$.
Note that these approximations are valid for the following two assumptions \cite{Sugiyama, Breuer2001}: the first assumption is that the de Brogile wavelength is smaller than the wavepacket width of particle.
The second assumption is that the Compton wavelength $\lambda_{\text{C}}$ of the charged particle is much shorter than the wavelength of photon field
$\lambda_\text{ph}$ (for example, the wavelength of photon field emitted from charged particle) ($\lambda_{\text{C}} \ll \lambda_\text{ph}$).
The initial state evolves as follows:
\begin{align}
|\Psi(T)\rangle 
&= 
\exp\big[-i \hat{H} T\big]|\Psi(0)\rangle
\nonumber\\
&=
e^{-i\hat{H}_0T} 
\text{T}\exp\big[-i\int_{0}^{T} dt \hat{V}_\text{I}(t) \big]
|\Psi(0)\rangle
\nonumber\\
&\approx 
e^{-i\hat{H}_0T}\frac{1}{2}
\sum_\text{P,Q=R,L}|\text{P}\rangle_{\text{A}} |\text{Q}\rangle_{\text{B}} \hat{U}_\text{PQ} |\alpha\rangle_\text{ph}
\nonumber 
\\
&=
\frac{1}{2}
\sum_\text{P,Q=R,L}|\text{P}_\text{f} \rangle_{\text{A}} |\text{Q}_\text{f} \rangle_{\text{B}} \, e^{-i\hat{H}_\text{ph} T} \hat{U}_\text{PQ} |\alpha\rangle_\text{ph},
\label{state}
\end{align}
where $T(>T_\text{A})$ is the total time scale while particle A is spatially superposed.
We used the approximations given by \eqref{approx3} in the third line. 
$|\text{P}_\text{f}\rangle_{\text{A}}=e^{-i\hat{H}_{\text{A}}T}|\text{P}\rangle_{\text{A}}$ and $|\text{Q}_\text{f}\rangle_{\text{B}}=e^{-i\hat{H}_{\text{B}}T}|\text{Q}\rangle_{\text{B}}$ with $\text{P}, \text{Q}=\text{R}, \text{L}$ are the states of charged particles A and B, which moved along the trajectories P and Q, respectively.
The unitary operator $\hat{U}_{\text{PQ}}$ is given by
\begin{align}
\hat{U}_{\text{PQ}} &=\text{T} \exp \left[-i \int_{0}^{T} d t \int d^{3} x\left(J_{\text{AP}}^{\mu}+J_{\text{BQ}}^{\mu}\right) \hat{A}_{\mu}^{\text{I}}(x)\right],
\label{unitaryPQ}
\end{align}
where 
$\text{T}$ denotes the time ordering, and 
$\hat{A}^\text{I}_\mu$ is the photon field operator in the interaction picture.
For convenience, we rewrite the state given in \eqref{state} as  
\begin{align}
|\Psi(T)\rangle
&=
\frac{1}{2}
\sum_\text{P,Q=R,L}|\text{P}_\text{f} \rangle_{\text{A}} |\text{Q}_\text{f} \rangle_{\text{B}} \, e^{-i\hat{H}_\text{ph} T} \hat{U}_\text{PQ} |\alpha\rangle_\text{ph}
\nonumber\\
\quad
&
=
\frac{1}{\sqrt{2}}
|\text{R}_\text{f}\rangle_{\text{A}}|\Omega_{\text{R}}\rangle_\text{B,ph}
+
\frac{1}{\sqrt{2}}
|\text{L}_\text{f}\rangle_{\text{A}}|\Omega_{\text{L}}\rangle_\text{B,ph},
\label{stateA}
\end{align}
where we defined
\begin{align}
|\Omega_{\text{P}}\rangle_\text{B,ph}
=
\frac{1}{\sqrt{2}}
\sum_{\text{Q}=\text{R},\text{L}}|\text{Q}_{\text{f}}\rangle_{\text{B}}
e^{-i\hat{H}_\text{ph} T}\hat{U}_{\text{PQ}}|\alpha\rangle_{\text{ph}}.
\end{align}
The vector 
$|\Omega_\text{P}\rangle_\text{B,ph}$ describes 
the composite state of particle B and the photon field when particle A moves along the trajectory P. 
The quantum state of particle A is obtained by tracing out the degrees of freedom of particle B and the photon field:
\begin{align}
\rho_{\text{A}}
&=\text{Tr}_{\text{B}, \text{ph}}[|\Psi(T)\rangle \langle\Psi(T)|]
\nonumber\\
\quad
&=
\frac{1}{2}
\begin{pmatrix}
1&\frac{1}{2}e^{-\Gamma_{\text{A}}+i\Phi_\text{A}}\Big(e^{-i\int d^4x(J^{\mu}_{\text{AR}}-J^{\mu}_{\text{AL}}) A_{\text{BR} \mu}}+e^{-i\int d^4x(J^{\mu}_{\text{AR}}-J^{\mu}_{\text{AL}})A_{\text{BL}\mu}}\Big)
\\
\quad * \quad &1
\end{pmatrix}
\label{rhoA},
\end{align}
where we used the basis 
$\{|\text{R}_\text{f} \rangle_\text{A}, |\text{L}_\text{f} \rangle_\text{A} \}$ to represent the density operator, and 
$*$ is the complex conjugate of the 
$(\text{R},\text{L})$ component. 
$A^{\mu}_\text{BQ} (\text{Q}=\text{R},\text{L})$ is the retarded photon field caused by charged particle B,
\begin{align}
A^{\mu}_{\text{BQ}}(x)=\int d^4y 
G^{\text{r},\mu} {}_{\nu} (x,y) J^\nu_\text{BQ} (y),
\label{retphotonB}
\end{align}
with the retarded Green's function, 
\begin{align}
G^\text{r}_{\mu\nu} (x,y)=-i[\hat{A}^\text{I}_\mu(x), \hat{A}^\text{I}_\nu (y) ]\theta(x^0-y^0). 
\end{align}
The quantities $\Gamma_\text{A}$ and $\Phi_\text{A}$ are 
\begin{align}
\Gamma_{\text{A}}
&
=
\frac{1}{4}\int d^4x d^4y\big(J^{\mu}_{\text{AR}}(x)-J^{\mu}_{\text{AL}}(x)\big)\big(J^{\mu}_{\text{AR}}(y)-J^{\mu}_{\text{AL}}(y)\big)
\langle\{\hat{A}^{\text{I}}_{\mu}(x), \hat{A}^{\text{I}}_{\mu}(y)\} \rangle,
\label{eq:GammaA}
\\
\quad
\Phi_\text{A}
&=
\int d^4x (J^{\mu}_{\text{AR}}(x)-J^{\mu}_{\text{AL}}(x))A_{\mu}(x)
-\frac{1}{2}\int d^4x d^4y(J^{\mu}_{\text{AR}}(x)-J^{\mu}_{\text{AL}}(x))(J^{\nu}_{\text{AR}}(y)+J^{\nu}_{\text{AL}}(y))G^{\text{r}}_{\mu\nu}(x,y),
\label{eq:PhiA}
\end{align}
where 
$\langle \cdot \rangle$ denotes the vacuum expectation value and 
$A_{\mu}(x)$ is defined in Appendix \ref{trace}.
The derivation of the density operator 
$\rho_\text{A}$ is presented in Appendix \ref{trace}.
The quantity $\Gamma_{\text{A}}$ characterizes the decoherence effect due to the radiation of the on-shell photon emitted by particle A \cite{Danielson2021,Sugiyama}.
The result \eqref{rhoA} with the retarded photon field $A^{\mu}_\text{BQ}$ of particle B implies that the effect of particle B can propagate to Alice's system.
However, in the spacelike case 
$D > T_\text{A}$ and $D > T_\text{B}$ (see Fig. \ref{fig:configuration}), the photon field induced by particle B does not reach particle A, i.e., $A^{\mu}_{\text{BQ}}(x)=0$.
Thus, the density operator \eqref{rhoA} becomes
\begin{align}
\rho_{\text{A}}
&=
\frac{1}{2}
\begin{pmatrix}
1&e^{-\Gamma_{\text{A}}+i\Phi_\text{A}}
\\
e^{-\Gamma_{\text{A}}-i\Phi_\text{A}}&1
\end{pmatrix}.
\end{align}
This result indicates that the process of charged particle B during the time $T_\text{B}$ does not affect the interference experiment on charged particle A by causality.
Note that, given the law of charge conservation, we also have to consider the contribution from charged particle B before the time $T_\text{B}$.
Even by considering this, we can see that the density operator $\rho_\text{A}$ does not depend on influences from spacelike separated regions.
In the derivation of the above equations, for simplicity, we only discussed the contribution from particle B during the time $T_\text{B}$.
In the next section, we confirm that the paradox does not appear from the viewpoints of \textit{visibility} and \textit{distinguishability}.

\section{consistency of complementarity\label{complementarity}}
In this section, we introduce the visibility $\mathcal{V}_{\text{A}}$ of charged particle A and the distinguishability $\mathcal{D}_{\text{B}}$ which quantifies the which-path information of particle A acquired through charged particle B. 
These two quantities are useful for expressing complementarity. 
Additionally, we discuss the relationship with the Robertson inequality in the last subsection.
According to Refs. \cite{Jaeger, Englert}, the visibility $\mathcal{V}_{\text{A}}$ and the distinguishability $\mathcal{D}_{\text{B}}$ satisfy the inequality, 
\begin{align}
\mathcal{V}^2_{\text{A}}+\mathcal{D}^2_{\text{B}} \leq 1.
\label{inequality}
\end{align}
This inequality expresses the complementarity: 
if the distinguishability is unity, $D_\text{B}=1$, the visibility $\mathcal{V}_\text{A}$ vanishes, and if the visibility is unity, $\mathcal{V}_\text{A}=1$, the distinguishability  $\mathcal{D}_\text{B}$ vanishes. 
In Appendix \ref{proof}, we present a simple proof of the above inequality by using the definitions of visibility and distinguishability described in the next Subsection A.
\subsection{Visibility and distinguishability}
We introduce the visibility
$\mathcal{V}_{\text{A}}$ of charged particle A defined as
\begin{align}
\mathcal{V}_{\text{A}}
=2|_{\text{A}}\langle \text{L}_{\text{f}}|\rho_{\text{A}}|\text{R}_{\text{f}}\rangle_{\text{A}}|,
\label{visibility2}
\end{align}
where $\rho_{\text{A}}$ is the reduced density operator of particle A given in Eq.~\eqref{rhoA}. 
The visibility $\mathcal{V}_{\text{A}}$ describes the extent to which the coherence of charged particle A remains when Alice performs an interference experiment.
Using Eq. \eqref{rhoA}, we have 
\begin{align}
\mathcal{V}_{\text{A}}
=e^{-\Gamma_{\text{A}}}
\left|
\cos
\Big(
\frac{\Phi_{\text{AB}}}{2}
\Big)
\right|,
\label{visibility}
\end{align}
where 
$\Phi_{\text{AB}} =\int d^4x(J^{\mu}_{\text{AR}}-J^{\mu}_{\text{AL}})\Delta A_{\text{B}\mu}$ with
$\Delta{A}^\mu_\text{B}=A^\mu_\text{BR}-A^\mu_\text{BL}$. 
For the case 
$D>T_\text{A}$ and 
$D>T_\text{B}$, the retarded photon field induced by charged particle B during time $T_\text{B}$ is zero ($A^\mu_\text{BQ}=0$, with  $\text{Q}=\text{R},\text{L}$). 
Then, the visibility is simply written as 
$\mathcal{V}_\text{A}=e^{-\Gamma_\text{A}}$ with 
$\Gamma_\text{A}$, which quantifies the decoherence effect due to the radiation of photon field emitted from particle A. 

Next, we introduce the distinguishability computed from the state of charged particle B. 
Tracing over particle A and the photon field from the state given in 
\eqref{state}, we obtain the state of particle B:
\begin{align}
\text{Tr}_{\text{A}, \text{ph}}[|\Psi(T)\rangle \langle\Psi(T)|]
&=
\frac{1}{2}
\text{Tr}_{\text{ph}}[|\Omega_\text{R}\rangle_\text{B,ph} \langle\Omega_\text{R}|]
+
\frac{1}{2}
\text{Tr}_\text{ph}[|\Omega_\text{L}\rangle_\text{B,ph} \langle \Omega_\text{L}|]
\nonumber\\
\quad
&
=
\frac{1}{2}\rho_{\text{BR}}+\frac{1}{2}\rho_{\text{BL}},
\label{rhoB}
\end{align}
where we defined $\rho_{\text{BP}}=\text{Tr}_{\text{ph}}[|\Omega_\text{P}\rangle_\text{B,ph} \langle\Omega_\text{P}|]$ with $\text{P}=\text{R}, \text{L}$ in the second line.
The density operator 
$\rho_\text{BP}$ describes the state of particle B when particle A moves along the trajectory P.  
The distinguishability $\mathcal{D}_{\text{B}}$ which characterizes how Bob can distinguish the trajectory of particle A from the state of particle B is defined as
\begin{align}
\mathcal{D}_{\text{B}}
=
\frac{1}{2}\text{Tr}_{\text{B}}|\rho_{\text{BR}}-\rho_{\text{BL}}|. 
\end{align}
where 
$\text{Tr}|\hat{O}|=\sum_i |\lambda_i|$ is given by the eigenvalues $\lambda_i$ 
of a Hermitian operator $\hat{O}$.
The distinguishability is nothing but the trace distance between the density operators $\rho_{\text{BR}}$ and $\rho_{\text{BL}}$ 
\cite{Nielsen}.
If the distinguishability vanishes, $\mathcal{D}_\text{B}=0$, and the two density operators 
$\rho_\text{BR}$ and 
$\rho_\text{BL}$ are identical. 
This means that Bob cannot know which trajectory particle A has taken from the state of particle B. 
However, 
if $\mathcal{D}_\text{B}=1$, 
the density operators 
$\rho_\text{BR}$ and 
$\rho_\text{BL}$ are orthogonal to each other ($\rho_\text{BR}\rho_\text{BL}=0$). 
Then, by measuring the state of particle B, Bob can guess which trajectory particle A has passed through. 
In this sense, the distinguishability $\mathcal{D}_\text{B}$ quantifies the amount of which-path information of particle A. 
The general property of the trace distance is presented in \cite{Nielsen}, and the meaning of the distinguishability mentioned above was discussed in \cite{Englert}.

Using the expression for the density operator 
$\rho_\text{BP}$ presented in Appendix \ref{trace}, we obtain the eigenvalues of the density operator $\rho_{\text{BR}}-\rho_{\text{BL}}$ as
\begin{align}
\lambda_{\pm}
&=\pm\frac{1}{2}\left|e^{-\Gamma_\text{B}+i\Phi_\text{B}-i\int d^4x(J^{\mu}_{\text{BR}}-J^{\mu}_{\text{BL}})A_{\text{R}\mu}}-e^{-\Gamma_\text{B}+i\Phi_\text{B}-i\int d^4x(J^{\mu}_{\text{BR}}-J^{\mu}_{\text{BL}})A_{\text{L}\mu}}\right|
\nonumber\\
&
=\pm e^{-\Gamma_{\text{B}}}\left|\sin\Big(\frac{1}{2}\int d^4x(J^{\mu}_{\text{BR}}-J^{\mu}_{\text{BL}})\Delta A_{\text{A}\mu}\Big)\right|,
\end{align}
where $\Delta A^{\mu}_{\text{A}}=A^{\mu}_{\text{AR}}-A^{\mu}_{\text{AL}}$ with
\begin{align}
A^{\mu}_{\text{AP}}(x)=\int d^4y G^{\text{r}, \mu}{}_{\nu}(x,y) J^\nu_\text{AP}(y), 
\label{retphotonA}
\end{align}
and 
$\Gamma_\text{B}$ and 
$\Phi_\text{B}$
are 
\begin{align}
\Gamma_{\text{B}}
&
=
\frac{1}{4}\int d^4x d^4y\big(J^{\mu}_{\text{BR}}(x)-J^{\mu}_{\text{BL}}(x)\big)\big(J^{\mu}_{\text{BR}}(y)-J^{\mu}_{\text{BL}}(y)\big)
\langle\{\hat{A}^{\text{I}}_{\mu}(x), \hat{A}^{\text{I}}_{\mu}(y)\} \rangle,
\label{eq:GammaB}
\\
\Phi_\text{B}
&=
\int d^4x (J^{\mu}_{\text{BR}}(x)-J^{\mu}_{\text{BL}}(x))A_{\mu}(x)
-\frac{1}{2}\int d^4x d^4y(J^{\mu}_{\text{BR}}(x)-J^{\mu}_{\text{BL}}(x))(J^{\nu}_{\text{BR}}(y)+J^{\nu}_{\text{BL}}(y))G^{\text{r}}_{\mu\nu}(x,y).
\label{LLLR}
\end{align}
The quantity $\Gamma_\text{B}$ characterizes the dephasing effect induced by the vacuum fluctuations of the photon field around particle B (see Subsection B or Refs. \cite{Sugiyama, Stern, Ford1997}).
The distinguishability is computed as
\begin{align}
\mathcal{D}_{\text{B}}=\frac{1}{2}(|\lambda_{+}|+|\lambda_{-}|)
=
e^{-\Gamma_{\text{B}}}\left|\sin\Big(\frac{\Phi_\text{BA}}{2}\Big)\right|,
\end{align}
where 
$\Phi_{\text{BA}} = \int d^4x(J^{\mu}_{\text{BR}}-J^{\mu}_{\text{BL}})\Delta A_{\text{A}\mu}$, and therefore, the inequality \eqref{inequality} is expressed as
\begin{align}
\mathcal{V}^2_{\text{A}}+\mathcal{D}^2_{\text{B}}
=
e^{-2\Gamma_{\text{A}}}\cos^2\left(\frac{\Phi_{\text{AB}}}{2}\right)
+
e^{-2\Gamma_{\text{B}}}\sin^2\left(\frac{\Phi_{\text{BA}}}{2}\right)
\leq 1. 
\end{align}
For the case 
$D>T_\text{A}$ and $D>T_\text{B}$, the retarded photon field of particle B vanishes ($A^{\mu}_{\text{BP}}=0$), which leads to $\Phi_{\text{AB}}=0$, and we have 
\begin{align}
\mathcal{V}^2_{\text{A}}+\mathcal{D}^2_{\text{B}}
=
e^{-2\Gamma_{\text{A}}}
+
e^{-2\Gamma_{\text{B}}}\sin^2\left(\frac{\Phi_{\text{BA}}}{2}\right)
\leq 1.
\label{fringein}
\end{align}
This inequality is consistent with the existence of the quantum radiation emitted from particle A ($\Gamma_\text{A}>0$) and the vacuum fluctuations of the photon field around particle B ($\Gamma_\text{B} > 0$) when the causality holds.
If we can remove the two effects 
($\Gamma_\text{A}=\Gamma_\text{B}=0$), this inequality would be violated as long as the retarded photon field of particle A does not vanish ($A^\mu_\text{AP} \neq 0$ and then
$\Phi_\text{BA} \neq 0$).
Hence, if the two effects vanish, then complementarity is violated, and the paradox would appear.
In the following subsection, we will discuss that the inequality \eqref{fringein} is never violated by the Robertson inequality associated with the photon field.

\subsection{Relationship with uncertainty relation}
In Refs. \cite{Sugiyama, Stern, Ford1997}, the quantity $\Gamma_{i}$ ($i=\text{A}, \text{B}$) was evaluated as the dephasing effect due to the vacuum fluctuations of the photon field,
\begin{align}
\langle 0| e^{i\hat{\phi}_{i}}|0\rangle
=
e^{-\langle 0|\hat{\phi}^2_i|0\rangle/2}
=
e^{-\Gamma_{i}},
\end{align}
with the operators $\hat{\phi}_{\text{A}}$ and $\hat{\phi}_{\text{B}}$ defined by 
\begin{align}
\hat{\phi}_{\text{A}}=\int d^4x (J^{\mu}_{\text{AR}}(x)-J^{\mu}_{\text{AL}}(x))\hat{A}^{\text{I}}_{\mu}(x)
,\quad
\hat{\phi}_{\text{B}}=\int d^4x (J^{\mu}_{\text{BR}}(x)-J^{\mu}_{\text{BL}}(x))\hat{A}^{\text{I}}_{\mu}(x),
\end{align}
where $\hat{A}^\text{I}_\mu$ is the photon field operator in the interaction picture, and $J^\mu_\text{AP}$ and
$J^\mu_\text{BQ}$ are the charged currents of each particle. 
The operators 
$\hat{\phi}_\text{A}$ and 
$\hat{\phi}_\text{B}$
describe the phase shifts due to the quantum fluctuations of the photon field.
The variances of $\hat{\phi}_{\text{A}}$ and $\hat{\phi}_{\text{B}}$ are related to the quantities $\Gamma_{\text{A}}$ and $\Gamma_{\text{B}}$ as follows:
\begin{align}
(\Delta {\phi}_{\text{A}})^2
=\langle0|\hat{\phi}^2_{\text{A}}|0\rangle-(\langle0|\hat{\phi}_{\text{A}}|0\rangle)^2
=2\Gamma_{\text{A}}
,\quad 
(\Delta {\phi}_{\text{B}})^2
=\langle0|\hat{\phi}^2_{\text{B}}|0\rangle-(\langle0|\hat{\phi}_{\text{B}}|0\rangle)^2
=2\Gamma_{\text{B}}.
\label{vacuum}
\end{align}
In the following equations, we show that the product of $\Gamma_{\text{A}}$ and $\Gamma_{\text{B}}$ has a lower bound given by the quantity $\Phi_\text{BA}$.
To observe this, we focus on the commutation relation of the operators $\hat{\phi}_{\text{A}}$ and $\hat{\phi}_{\text{B}}$,
\begin{align}
[\hat{\phi}_{\text{A}}, \hat{\phi}_{\text{B}}]
&=
\int d^4x d^4y (J^{\mu}_{\text{AR}}(x)-J^{\mu}_{\text{AL}}(x))(J^{\nu}_{\text{BR}}-J^{\nu}_{\text{BL}}(y))[\hat{A}^{\text{I}}_{\mu}(x), \hat{A}^{\text{I}}_{\nu}(y)]
\nonumber\\
\quad
&=
\int d^4x d^4y (J^{\mu}_{\text{AR}}(x)-J^{\mu}_{\text{AL}}(x))(J^{\nu}_{\text{BR}}-J^{\nu}_{\text{BL}}(y))[\hat{A}^{\text{I}}_{\mu}(x), \hat{A}^{\text{I}}_{\nu}(y)]\theta (x^{0}-y^{0})
\nonumber\\
\quad
&+
\int d^4x d^4y (J^{\mu}_{\text{AR}}(x)-J^{\mu}_{\text{AL}}(x))(J^{\nu}_{\text{BR}}-J^{\nu}_{\text{BL}}(y))[\hat{A}^{\text{I}}_{\mu}(x), \hat{A}^{\text{I}}_{\nu}(y)]\theta (y^{0}-x^{0})
\nonumber\\
\quad
&=
i\int d^4x(J^{\mu}_{\text{AR}}-J^{\mu}_{\text{AL}})\Delta A_{\text{B}\mu}-i\int d^4x(J^{\mu}_{\text{BR}}-J^{\mu}_{\text{BL}})\Delta A_{\text{A}\mu}
\nonumber\\
\quad
&=
-i\Phi_{\text{BA}},
\end{align}
where we inserted the step functions $\theta (x^{0}-y^{0})+\theta (y^{0}-x^{0})$ in the second line, and we changed variables as $x^{\mu} \leftrightarrow y^{\mu}$ and indices as ${\mu} \leftrightarrow {\nu}$ of the second term in the third line.
Note that the first term $i\Phi_{\text{AB}}=i\int d^4x(J^{\mu}_{\text{AR}}-J^{\mu}_{\text{AL}})\Delta A_{\text{B}\mu}$ in the third line vanished by assuming the case 
$D>T_\text{A}$ and 
$D>T_\text{B}$, where there is no retarded propagation of photon field from Bob's system to Alice's system.
This commutation relation shows that the operators $\hat{\phi}_{\text{A}}$ and $\hat{\phi}_{\text{B}}$ do not commute with each other because 
the influence of particle A causally propagates to particle B from the far past (the red or blue line in Fig. \ref{fig:configuration}) and then $\Phi_\text{BA} \neq 0$.
Using this commutation relation, we obtain the following Robertson inequality as
\begin{align}
(\Delta {\phi}_{\text{A}})^2(\Delta {\phi}_{\text{B}})^2
\geq 
\frac{1}{4}\left|\langle0|[\hat{\phi}_{\text{A}}, \hat{\phi}_{\text{B}}]|0\rangle\right|^2=\frac{1}{4}\Phi^2_{\text{BA}}.
\end{align}
From \eqref{vacuum}, we get the inequality among $\Gamma_{\text{A}}$, $\Gamma_{\text{B}}$ and $\Phi_{\text{BA}}$,
\begin{align}
\Gamma_{\text{A}}\Gamma_{\text{B}}
\geq
\frac{1}{16}\Phi^2_{\text{BA}}.
\label{robert}
\end{align}
This means that the quantities 
$\Gamma_\text{A}$ and 
$\Gamma_\text{B}$ do not vanish simultaneously if 
$\Phi_\text{BA}\neq0$. 
Additionally, we can show that the Robertson inequality \eqref{robert} is a sufficient condition for the inequality \eqref{fringein}:
\begin{align}
\Gamma_{\text{A}}\Gamma_{\text{B}}
\geq
\frac{1}{16}\Phi^2_{\text{BA}}
\quad
\Longrightarrow
\quad
e^{-2\Gamma_{\text{A}}}+e^{-2\Gamma_{\text{B}}}\sin^2\left(\frac{\Phi_{\text{BA}}}{2}\right)
\leq 1.
\label{suffcon}
\end{align}
The proof of this statement is presented in Appendix \ref{proofin}.
This result implies that the Robertson inequality among $\Gamma_{\text{A}}$, $\Gamma_{\text{B}}$ and $\Phi_{\text{BA}}$, which reflects the non-commutative property of the photon field, guarantees the complementarity described by the inequality between the visibility $\mathcal{V}_\text{A}$ and the distinguishability 
$\mathcal{D}_\text{B}$.

\section{conclusion\label{conclusion}}
In this study, we revisited
the resolution of the paradox proposed by Belenchia et al. \cite{Belenchia2018, Belenchia2019} in the system of a photon field interacting with two charged particles in the superposition states of two trajectories.
The analysis based on the quantum field theory explicitly demonstrated the intuitively legitimate result that causality holds and that operations on Bob's system at a spacelike distance do not affect Alice's interference experiment at all by deriving Alice's reduced density operator.
On the other hand, to find the validity of complementarity, we first derived visibility and distinguishability, which represent the degree of success of Alice's interference experiment and the degree of distinction of Bob's quantum state, respectively.
Then, we argued that there is an inequality between these quantities, which is guaranteed by the Robertson inequality associated with the non-commutative property of the photon field (the quantized electromagnetic field).
This inequality describes the limit of complementarity in resolving this paradox. 

Thus, to resolve this paradox, the fact that the photon field has a non-commutative 
property is the most important factor in our analysis.
This conclusion is applicable to gravitational interactions.  
A similar analysis of the gravitational version of the present paper should be 
performed explicitly in future work, but the results will be inferred with reference to our analysis, as follows. 
Let us consider the massive particles A and B.
According to the analogy in Section \ref{complementarity}, the phase shifts induced by the quantum fluctuations of gravitational field can be described as follows:
\begin{align}
\hat{\phi}^{\text{g}}_{\text{A}}=\int d^4x (T^{\mu \nu}_{\text{AR}}(x)-T^{\mu \nu}_{\text{AL}}(x))\hat{h}^{\text{I}}_{\mu \nu}(x)
,\quad
\hat{\phi}^{\text{g}}_{\text{B}}=\int d^4x (T^{\mu \nu}_{\text{BR}}(x)-T^{\mu \nu}_{\text{BL}}(x))\hat{h}^{\text{I}}_{\mu \nu}(x),
\end{align}
where $\hat{h}^{\text{I}}_{\mu \nu}$ is the linearized quantum gravitational field in the interaction picture which is the perturbation from the Minkowski spacetime, and $T^{\mu \nu}_{i\text{P}}$ ($i=\text{A}, \text{B}$ and $\text{P}=\text{R}, \text{L}$) is the energy-momentum tensor of each massive particle. 
Hence the decoherence (dephasing) effects due to the vacuum fluctuations can be characterized by
\begin{align}
\Gamma^{\text{g}}_{\text{A}}=\frac{1}{2}
\langle0|
(\hat{\phi}^{\text{g}}_{\text{A}})^2
|0\rangle
,\quad
\Gamma^{\text{g}}_{\text{B}}=\frac{1}{2}
\langle0|
(\hat{\phi}^{\text{g}}_{\text{B}})^2
|0\rangle,
\end{align}
and are limited by the phase shift induced by the retarded gravitational field owing to the Robertson inequality:
\begin{align}
\Gamma^{\text{g}}_{\text{A}}\Gamma^{\text{g}}_{\text{B}}
\geq
\frac{1}{16}{(\Phi^{\text{g}}_{\text{BA}})^2},
\end{align}
where $\Phi^{\text{g}}_{\text{BA}}$ is defined by
\begin{align}
\Phi^{\text{g}}_{\text{BA}} 
&\equiv
\int d^4x(T^{\mu \nu}_{\text{BR}}-T^{\mu \nu}_{\text{BL}})\Delta h^{\text{A}}_{\mu \nu},
\end{align}
with the retarded gravitational field, 
\begin{align}
\Delta h^{\text{A}}_{\mu \nu}(x)=\int d^4y (T^{\rho\sigma}_{\text{AR}}(y)-T^{\rho \sigma}_{\text{AL}}(y))G^{\text{r}}_{\mu\nu\rho\sigma}(x,y).
\end{align}
Note that the function $G^{\text{r}}_{\mu\nu\rho\sigma}(x,y)$ is the retarded Green's function, and the detailed formula is given in \cite{Marshman, Rivers}.
In the gravitational version of our analysis, the consistency between causality and complementarity is guaranteed by the Robertson inequality.
Repeating the discussion of Belenchia et al. \cite{Belenchia2018, Belenchia2019,Danielson2021}, we suggest that the quantities  $\Gamma^{\text{g}}_{\text{A}}$ and $\Gamma^{\text{g}}_{\text{B}}$ do not vanish at the same time so that either $\Gamma^{\text{g}}_{\text{A}}$ or $\Gamma^{\text{g}}_{\text{B}}$ must be caused by the on-shell gravitational radiation from Alice's particle A and the vacuum fluctuation of the gravitational field around Bob's particle B.
This shows the necessity of the non-commutative property of the gravitational field related to the Robertson inequality.

\textit{Note added:}
Recently the authors of Ref.~\cite{Iso} revisited the same paradox by assuming a simple theoretical model so that Alice with a spin and Bob with a continuous variable are 
coupled to each other through a quantized scalar field. 
They focused on the quantity $\langle \Psi_{\downarrow}|\Psi_{\uparrow}\rangle_{\phi, B}=e^{-\gamma_{\text{A}}}\delta_{\epsilon}(M)$, which denotes the interference term of Alice's state after tracing out the states of the scalar field $\phi$ and Bob's states.
The quantity $e^{-\gamma_{\text{A}}}$ represents the decoherence due to the vacuum fluctuations of the scalar field $\phi$, while $\delta_{\epsilon}(M)$ is an overlap of 
the wave function of Bob's system with $M$ described with the retarded Green's function propagating from Bob to Alice. 
Therefore, $\gamma_A$ and $M$ in their study \cite{Iso} correspond to $\Gamma_{\text{A}}$ and $\Phi_{\text{AB}}$, respectively.
Therefore, $e^{-\gamma_{\text{A}}}\delta_{\epsilon}(M)$ corresponds to the visibility function \eqref{visibility} in the present study. 
The primary purpose of our study in the present paper is to demonstrate that the consistency between causality and complementarity is guaranteed by the Robertson inequality of the quantized field in the positions of Alice and Bob.
This reflects the existence of a gravitational field with quantum non-commutativity.

\acknowledgements
We are grateful for the discussions at the QUP theoretical collaboration.
We especially thank S. Iso, Y. Hidaka,  J. Soda, Y. Nambu, K. Shimada, and Y. Kuramochi 
for their insightful discussions and helpful comments.
Y.S. was supported by the Kyushu University Innovator Fellowship in Quantum Science.
A.M. was supported by 2022 Research Start Program 202203.
K.Y. was partially supported by JSPS KAKENHI, Grant No. 22H05263.

\begin{appendix}
\section{Derivation of 
the density operators 
$\rho_\text{A}$ and 
$\rho_\text{BP}$
\label{trace}}
In this Appendix, we derive the expression of the density operators $\rho_\text{A}$ and 
$\rho_\text{BP}$. 
To do this, we compute  $\Tr_{\text{ph}}[|\Omega_{\text{P}}\rangle_\text{B,ph} \langle \Omega_{\text{P}'}|]$ as follows:
\begin{align}
\Tr_{\text{ph}}[|\Omega_\text{P}\rangle_\text{B,ph}\langle \Omega_{\text{P}'}|]
&=
\frac{1}{2}\sum_{\text{Q}, \text{Q}'=\text{R}, \text{L}}
|\text{Q}_{\text{f}}\rangle_{\text{B}} \langle \text{Q}'_{\text{f}}|
{}_{\text{ph}}\langle \alpha |\hat{U}^{\dagger}_{\text{P}'\text{Q}'}\hat{U}_{\text{PQ}}|\alpha \rangle_\text{ph}
\nonumber\\
\quad
&=
\frac{1}{2}\sum_{\text{Q}, \text{Q}'=\text{R}, \text{L}}
e^{-\Gamma_{\text{P}'\text{Q}'\text{PQ}}+i\Phi_{\text{P}'\text{Q}'\text{PQ}}}
|\text{Q}_{\text{f}}\rangle_{\text{B}} \langle \text{Q}'_{\text{f}}|,
\end{align}
where 
${}_{\text{ph}}\langle \alpha |\hat{U}^{\dagger}_{\text{P}'\text{Q}'}\hat{U}_{\text{PQ}}|\alpha \rangle_\text{ph}=e^{-\Gamma_{\text{P}'\text{Q}'\text{PQ}}+i\Phi_{\text{P}'\text{Q}'\text{PQ}}}$ with the quantities, 
\begin{align}
\Gamma_{\text{P}'\text{Q}'\text{PQ}}&=\frac{1}{4}\int d^4x \int d^4y (J^\mu_{\text{P}'\text{Q}'}(x)-J^\mu_{\text{PQ}}(x))
(J^\nu_{\text{P}'\text{Q}'}(y)-J^\nu_{\text{PQ}}(y))
\langle \bigl\{\hat{A}^\text{I}_\mu (x), \hat{A}^\text{I}_\nu (y)\bigr\}\rangle,
\label{GammaP'Q'PQ}
\\
\Phi_{\text{P}'\text{Q}'\text{PQ}}
&=
\int d^4x (J^\mu_{\text{P}'\text{Q}'}(x)-J^\mu_{\text{PQ}}(x))A_\mu (x)-\frac{1}{2}\int d^4x \int d^4y (J^\mu_{\text{P}'\text{Q}'}(x)-J^\mu_{\text{PQ}}(x))(J^\nu_{\text{P}'\text{Q}'}(y)+J^\nu_{\text{PQ}}(y))G^\text{r}_{\mu \nu} (x,y),
\label{PhiP'Q'PQ}
\end{align}
was obtained in Appendix in \cite{Sugiyama}. 
$J^\mu_\text{PQ}=J^\mu_\text{AP}+J^\mu_\text{BQ}$ is given by the currents 
$J^\mu_\text{AP}$ and $J^\mu_\text{BQ}$ of charged particles A and B, respectively. 
The field $A_\mu (x)$ in \eqref{PhiP'Q'PQ} is the coherent photon field defined as
\begin{equation}
 A_\mu (x)= \int \frac{d^3k}{(2\pi)^{3/2} \sqrt{2k^0}}( \alpha_\mu (\bm{k})e^{ik_\nu x^\nu}+ c.c.),
 \label{A}
\end{equation}
and the complex function $\alpha_{\mu}(\bm{k})$ satisfies 
\begin{equation}
k^{\mu} \alpha_{\mu}(\bm{k})=-\frac{\tilde{J}^{0}(\bm{k})}{\sqrt{2k^{0}}}
\end{equation}
to guarantee the  Becchi-Rouet-Stora-Tyutin (BRST) condition (Appendix in \cite{Sugiyama}). 
Note that $\tilde{J}^0(\bm{k})=\tilde{J}^0_\text{A}(\bm{k})+\tilde{J}^0_\text{B}(\bm{k})$
is the eigenvalue of the Fourier transform of the charged current $\hat{\tilde{J}}^0(\bm{k})=\hat{\tilde{J}}^0_\text{A}(\bm{k})+\hat{\tilde{J}}^0_\text{B}(\bm{k})$ at the initial time 
$t=0$.
The function $\langle\{\hat{A}^{\text{I}}_{\mu}(x), \hat{A}^{\text{I}}_{\mu}(y)\} \rangle$ is the two-point function of the vacuum. 
We can compute the reduced density operator 
$\rho_\text{A}$ of the particle A in the basis $\{|\text{R}_\text{f} \rangle_\text{A}, |\text{L}_\text{f} \rangle_\text{A} \}$ as
\begin{align}
\rho_\text{A}
&=\text{Tr}_{\text{B},\text{ph}}[|\Psi(T)\rangle \langle \Psi(T)|]
\nonumber 
\\
&=\frac{1}{2} \sum_{\text{P},\text{P}'=\text{R},\text{L}} 
{}_\text{B,ph} \langle \Omega_{\text{P}'}|\Omega_\text{P} \rangle_\text{B,ph} 
|\text{P}_\text{f} \rangle_\text{A} \langle \text{P}'_\text{f}|
\nonumber 
\\
&=\frac{1}{4}  \sum_{\text{P},\text{P}'=\text{R},\text{L}} \sum_{\text{Q}=\text{R},\text{L}}
e^{-\Gamma_{\text{P}'\text{Q}\text{PQ}}+i\Phi_{\text{P}'\text{Q}\text{PQ}}}
|\text{P}_\text{f} \rangle_\text{A} \langle \text{P}'_\text{f}|
\nonumber 
\\
&=
\frac{1}{2}
\begin{pmatrix}
1&\frac{1}{2}\Big(e^{-\Gamma_\text{RRLR}+i\Phi_\text{RRLR}} +e^{-\Gamma_\text{RLLL}+i\Phi_\text{RLLL}}
\Big)
\\
\quad * \quad &1
\end{pmatrix}
\nonumber 
\\
&=
\frac{1}{2}
\begin{pmatrix}
1&\frac{1}{2}e^{-\Gamma_{\text{A}}+i\Phi_\text{A}}\Big(e^{-i\int d^4x(J^{\mu}_{\text{AR}}-J^{\mu}_{\text{AL}}) A_{\text{BR} \mu}}+e^{-i\int d^4x(J^{\mu}_{\text{AR}}-J^{\mu}_{\text{AL}})A_{\text{BL}\mu}}\Big)
\\
\quad * \quad &1
\end{pmatrix},
\end{align}
where 
${}_\text{B,ph}\langle \Omega_{\text{P}'}|\Omega_\text{P} \rangle_\text{B,ph}=\text{Tr}_\text{B}\big[\text{Tr}_\text{ph}[|\Omega_\text{P} \rangle_\text{B,ph} \langle \Omega_{\text{P}'}|]\big] $, and 
$*$ is the complex conjugate of the $(\text{R},\text{L})$ component.
Note that 
\begin{align}
\Gamma_\text{RRLR}&=\Gamma_\text{RLLL}=\Gamma_\text{A},
\\
\Phi_\text{RRLR}&=\Phi_\text{A}-\int d^4x(J^\mu_\text{AR}(x)-J^\mu_\text{AL}(x))A_{\text{BR}\mu}(x),
\\
\Phi_\text{RLLL}&=\Phi_\text{A}-\int d^4x(J^\mu_\text{AR}(x)-J^\mu_\text{AL}(x))A_{\text{BL}\mu}(x).
\end{align}
Here, 
$\Gamma_\text{A}$ and 
$\Phi_\text{A}$ are defined by Eqs.\eqref{eq:GammaA} and \eqref{eq:PhiA}, and the retarded field 
$A^\mu_\text{BQ}$ is given in \eqref{retphotonB}. 
The reduced density operators 
$\rho_\text{BR}$ and $\rho_\text{BL}$ in the basis 
$\{|\text{R}_\text{f}\rangle_\text{B}, |\text{L}_\text{f} \rangle_\text{B} \}$
are given as 
\begin{align}
\rho_\text{BR}
&=\text{Tr}_\text{ph}[|\Omega_\text{R}\rangle_\text{B,ph} \langle \Omega_\text{R}|]
\nonumber 
\\
&=
\frac{1}{2}\sum_{\text{Q}, \text{Q}'=\text{R}, \text{L}}
e^{-\Gamma_{\text{R}\text{Q}'\text{RQ}}+i\Phi_{\text{R}\text{Q}'\text{RQ}}}
|\text{Q}_{\text{f}}\rangle_{\text{B}} \langle \text{Q}'_{\text{f}}|,
\nonumber 
\\
&=
\frac{1}{2}
\begin{pmatrix}
1&e^{-\Gamma_\text{RRRL}+i\Phi_\text{RRRL}}
\\
\quad * \quad &1
\end{pmatrix}
\nonumber 
\\
&=
\frac{1}{2}
\begin{pmatrix}
1&e^{-\Gamma_\text{B}+i\Phi_\text{B}-i\int d^4x (J^\mu_\text{BR}-J^\mu_\text{BL})A_{\text{R}\mu}}
\\
\quad * \quad &1
\end{pmatrix},
\end{align}
and 
\begin{align}
\rho_\text{BL}
&=\text{Tr}_\text{ph}[|\Omega_\text{L}\rangle_\text{B,ph} \langle \Omega_\text{L}|]
\nonumber 
\\
&=
\frac{1}{2}\sum_{\text{Q}, \text{Q}'=\text{R}, \text{L}}
e^{-\Gamma_{\text{L}\text{Q}'\text{LQ}}+i\Phi_{\text{L}\text{Q}'\text{LQ}}}
|\text{Q}_{\text{f}}\rangle_{\text{B}} \langle \text{Q}'_{\text{f}}|,
\nonumber 
\\
&=
\frac{1}{2}
\begin{pmatrix}
1&e^{-\Gamma_\text{LRLL}+i\Phi_\text{LRLL}}
\\
\quad * \quad &1
\end{pmatrix}
\nonumber 
\\
&=
\frac{1}{2}
\begin{pmatrix}
1&e^{-\Gamma_\text{B}+i\Phi_\text{B}-i\int d^4x (J^\mu_\text{BR}-J^\mu_\text{BL})A_{\text{L}\mu}}
\\
\quad * \quad &1
\end{pmatrix},
\end{align}
where we used
\begin{align}
\Gamma_\text{RRRL}&=\Gamma_\text{LRLL}=\Gamma_\text{B},
\\
\Phi_\text{RRRL}&=\Phi_\text{B}-\int d^4x(J^\mu_\text{BR}(x)-J^\mu_\text{BL}(x))A_{\text{AR}\mu}(x),
\\
\Phi_\text{LRLL}&=\Phi_\text{B}-\int d^4x(J^\mu_\text{BR}(x)-J^\mu_\text{BL}(x))A_{\text{AL}\mu}(x),
\end{align}
where 
$\Gamma_\text{B}$ and 
$\Phi_\text{B}$ are defined in Eqs.\eqref{eq:GammaB} and \eqref{LLLR}, respectively.
The retarded field 
$A^\mu_\text{PQ}$ is given in \eqref{retphotonA}. 

\section{Proof of the inequality between visibility and distinguishability\label{proof}}
We prove the inequality  \eqref{inequality} between visibility and distinguishability.
First, we derive the visibility for the state given in \eqref{stateA}.
The visibility of charged particle A is calculated as
\begin{align}
\mathcal{V}_{\text{A}}
&
=
2|_{\text{A}}\langle \text{L}_{\text{f}}|\rho_{\text{A}}|\text{R}_{\text{f}}\rangle_{\text{A}}|
\nonumber\\
\quad
&
=
2|\text{Tr}_{\text{B}, \text{ph}}[_{\text{A}}\langle \text{L}_{\text{f}}|\Psi(T)\rangle \langle \Psi(T)|\text{R}_{\text{f}}\rangle_{\text{A}}]|
\nonumber\\
\quad
&
=
|{}_\text{B,ph} \langle\Omega_\text{R}|\Omega_\text{L}\rangle_\text{B,ph}|
\equiv
|\alpha|.
\label{overlap}
\end{align}
We next evaluate the distinguishability of charged particle B.
For a trace distance $\mathcal{D}(\rho, \sigma)$ with arbitrary density operators $\rho$ and $\sigma$, we use the fact that the trace-preserving quantum operations are contractive \cite{Nielsen}:
\begin{align}
\mathcal{D}(\mathcal{E}(\rho), \mathcal{E}(\sigma)) \leq \mathcal{D}(\rho, \sigma),
\label{monotonicity}
\end{align}
where $\mathcal{E}$ is a trace-preserving quantum operation.
This inequality means that the operation $\mathcal{E}$ makes it difficult to distinguish between the two quantum states $\rho$ and $\sigma$, i.e., the trace distance does not increase.
Then, the distinguishability is bounded as
\begin{align}
\mathcal{D}_{\text{B}}
&=
\frac{1}{2}\text{Tr}_{\text{B}}|\rho_{\text{BR}}-\rho_{\text{BL}}|
\nonumber\\
\quad
&=
\frac{1}{2}\text{Tr}_{\text{B}}|\text{Tr}_{\text{ph}}[|\Omega_\text{R}\rangle _\text{B,ph}\langle\Omega_{\text{R}}|]-\text{Tr}_{\text{ph}}[|\Omega_{\text{L}}\rangle_\text{B,ph} \langle\Omega_\text{L}|]|
\nonumber\\
\quad
&
\leq
\frac{1}{2}\text{Tr}_{\text{B}}||\Omega_\text{R}\rangle _\text{B,ph}\langle\Omega_{\text{R}}|-|\Omega_{\text{L}}\rangle_\text{B,ph} \langle\Omega_\text{L}||,
\end{align}
where the inequality \eqref{monotonicity} was used in the third line because the partial trace is a trace-preserving quantum operation.
To obtain the eigenvalues of the operator $|\Omega_{\text{R}}\rangle_\text{B,ph} \langle\Omega_\text{R}|-|\Omega_{\text{L}}\rangle_\text{B,ph} \langle\Omega_\text{L}|$, we define the orthonormal basis $\{|u_{\text{A}}\rangle, |u_{\text{B}}\rangle\}$ using the Gram-Schmidt orthonormalization as:
\begin{align}
|u_{\text{A}}\rangle=|\Omega_\text{R}\rangle_\text{B,ph}, \quad |u_{\text{B}}\rangle=\frac{|\Omega_\text{L}\rangle_\text{B,ph}-\alpha|\Omega_\text{R}\rangle_\text{B,ph}}{\sqrt{1-|\alpha|^2}},
\end{align}
where the overlap $\alpha$ is defined in \eqref{overlap}.
In this basis, the operator $|\Omega_{\text{R}}\rangle_\text{B,ph} \langle\Omega_\text{R}|-|\Omega_{\text{L}}\rangle_\text{B,ph} \langle\Omega_\text{L}|$ can be rewritten as
\begin{align}
|\Omega_{\text{R}}\rangle_\text{B,ph} \langle\Omega_\text{R}|-|\Omega_{\text{L}}\rangle_\text{B,ph} \langle\Omega_\text{L}|
&=|u_{\text{A}}\rangle \langle u_{\text{A}}|-(\alpha|u_{\text{A}}\rangle+\sqrt{1-|\alpha|^2}|u_{\text{B}}\rangle)(\alpha^{*}\langle u_{\text{A}}|+\sqrt{1-|\alpha|^2}\langle u_{\text{B}}|)
\nonumber\\
\quad
&=
\begin{pmatrix}
1-|\alpha|^2&\alpha\sqrt{1-|\alpha|^2}\\
\alpha^{*}\sqrt{1-|\alpha|^2}&-(1-|\alpha|^2)
\end{pmatrix},
\end{align}
in the orthonormal basis $\{|u_{\text{A}}\rangle, |u_{\text{B}}\rangle\}$.
Thus, the eigenvalues of this matrix $\lambda_{\text{A}, \text{B}}$ are
\begin{align}
\lambda_{\text{A}}=\sqrt{1-|\alpha|^2},
\quad
\lambda_{\text{B}}=-\sqrt{1-|\alpha|^2},
\end{align}
and the distinguishability $\mathcal{D}_{\text{B}}$ is suppressed by the sum of these eigenvalues as follows: 
\begin{align}
\mathcal{D}_{\text{B}}
\leq
\frac{1}{2}\text{Tr}_{\text{B}}||\Omega_{\text{R}}\rangle_\text{B,ph} \langle\Omega_\text{R}|-|\Omega_{\text{L}}\rangle_\text{B,ph} \langle\Omega_\text{L}||
=\frac{1}{2}(|\lambda_{\text{A}}|+|\lambda_{\text{B}}|)=\sqrt{1-|\alpha|^2}.
\label{proofdistin}
\end{align}
Substituting \eqref{overlap} into \eqref{proofdistin}, we find the relationship
\begin{align}
\mathcal{V}^2_{\text{A}}+\mathcal{D}^2_{\text{B}}
\leq
1.
\end{align}
Therefore, the visibility of charged particle A and the distinguishability of charged particle B follow the inequality \eqref{inequality}. 

\section{Proof of the statement in \eqref{suffcon} \label{proofin}}
We first numerically prove the statement in \eqref{suffcon}.
Using the Robertson inequality\eqref{robert}, 
$\Gamma_\text{A} \Gamma_\text{B} \geq \Phi^2_\text{AB}/16$, we have
\begin{align}
1-e^{-2\Gamma_{\text{A}}}-e^{-2\Gamma_{\text{B}}}\sin^2\left(\frac{\Phi_{\text{BA}}}{2}\right)
&
\geq
1-e^{-2\Gamma_{\text{A}}}-e^{-{\Phi^2_{\text{BA}}}/{8\Gamma_{\text{A}}}}\sin^2\left(\frac{\Phi_{\text{BA}}}{2}\right)=f(X,Y),
\end{align}
where we defined the function $f(X, Y)$ with 
$X = e^{-2\Gamma_{\text{A}}}$ and $Y = e^{-{\Phi^2_{\text{BA}}}/{8\Gamma_{\text{A}}}}$
as follows:
\begin{align}
f(X, Y)
=
1-X-Y\sin^2\left(\sqrt{\log X \log Y}\right).
\end{align}
As it is sufficient to consider that $\Gamma_{\text{A}}>0$ and 
$\Phi_{\text{BA}} > 0$, we can assume that $0< X <1$ and $0< Y <1$.
\begin{figure}[H]
  \centering
  \includegraphics[width=0.6\linewidth]{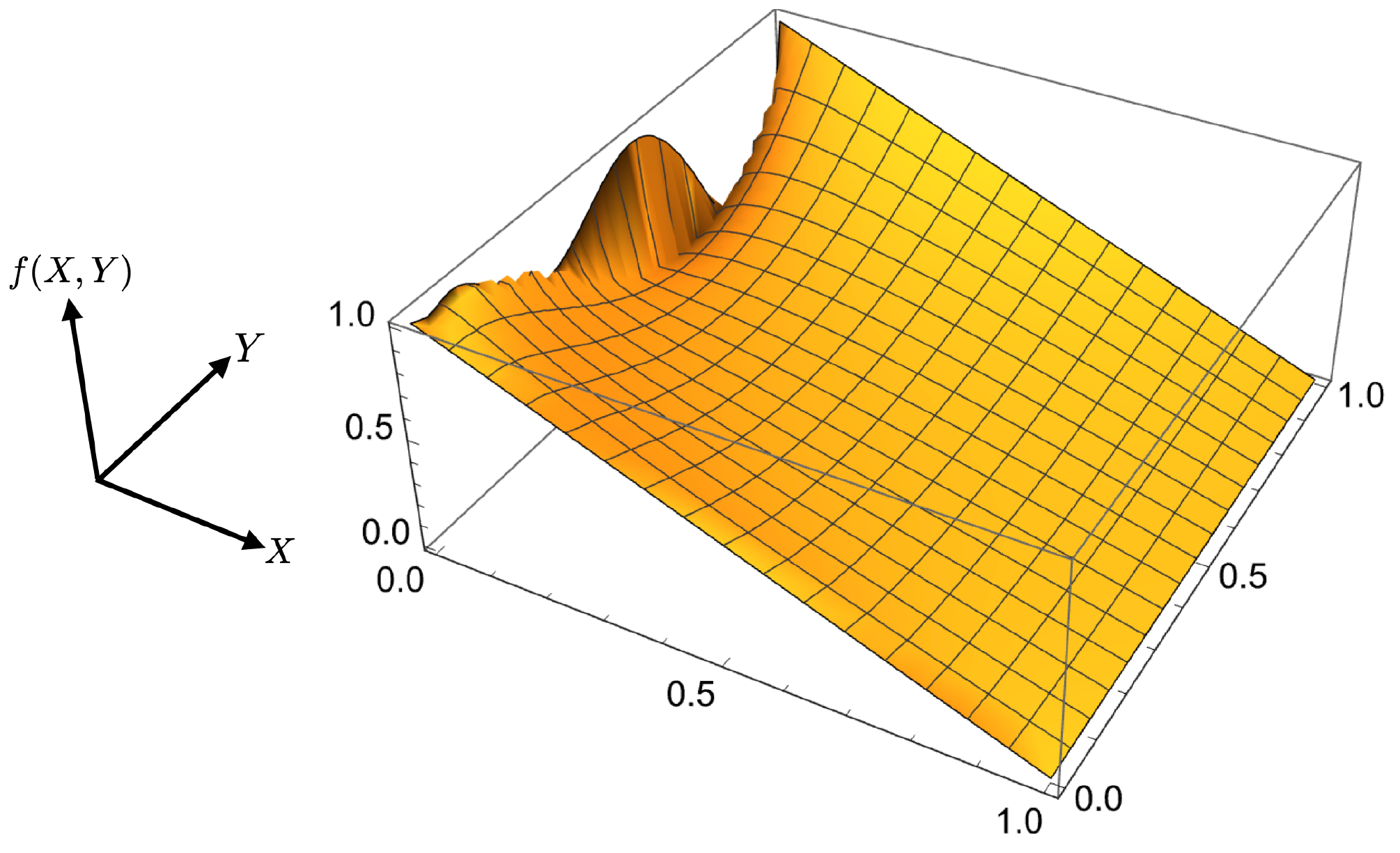}
  \caption{Behavior of the function $f(X, Y)$ where the region $0 < X < 1$ and $0< Y< 1$.}
  \label{fxy}
\end{figure}
\noindent
Fig. \ref{fxy} shows the behavior of the function $f(X, Y)$, which is positive in the regions $0 < X < 1$ and $0< Y< 1$.
Since the function 
$f(X,Y)$ is positive, the inequality $e^{-2\Gamma_{\text{A}}}+e^{-2\Gamma_{\text{B}}}\sin^2\left(\Phi_{\text{BA}}/2\right) \leq 1$ in \eqref{fringein} is satisfied. 
Hence, the Robertson inequality \eqref{robert} is the sufficient condition for the inequality \eqref{fringein}, and the statement in \eqref{suffcon} holds. 
In the following, we show that the function $f(X, Y)$ is always positive in an analytic manner.
\\
\noindent
\textit{Proof.}
Now let derive the partial derivatives to find the gradient for $f(X, Y)$, and the results are
\begin{align}
\frac{\partial f(X, Y)}{\partial X}
&=
-1
-\frac{Y\log Y \sin \left(\sqrt{\log X \log Y}\right)\cos \left(\sqrt{\log X \log Y}\right)}
{X\sqrt{\log X \log Y}}
,
\\
\frac{\partial f(X, Y)}{\partial Y}
&=
-
\left(
\frac{\log X \cos \left(\sqrt{\log X \log Y}\right)}{\sqrt{\log X \log Y}}
+
\sin \left(\sqrt{\log X \log Y}\right)
\right)
\sin \left(\sqrt{\log X \log Y}\right)
.
\end{align}
We are looking for the gradient is zero:
\begin{align}
0
=
\log X \cos \left(\sqrt{\log X \log Y}\right) 
+
\sqrt{\log X \log Y}\sin \left(\sqrt{\log X \log Y}\right),
\label{deriy}
\end{align}
and
\begin{align}
0
&=
-X\sqrt{\log X \log Y}-Y\log Y \sin \left(\sqrt{\log X \log Y}\right)\cos \left(\sqrt{\log X \log Y}\right)
\nonumber\\
\quad
&=
-X\left(\log X \log Y \right)-Y\log Y \left(\left(\sqrt{\log X \log Y}\right) \sin \left(\sqrt{\log X \log Y}\right)\right)\cos \left(\sqrt{\log X \log Y}\right),
\label{derix}
\end{align}
where we multiplied by the factor $\sqrt{\log X \log Y}$ in the second line.
Substituting \eqref{deriy} into \eqref{derix}, we obtain the following condition
\begin{align}
0
=
\left(\log X \log Y\right)\left(-X-Y\sin^2\left(\sqrt{\log X \log Y}\right)+Y\right).
\end{align}
\\
\noindent
Case 1: $\log X \log Y=0$, i.e., $X=1$ or $Y=1$.
When $X=1$, by definition of the function $f(X, Y)$, we have
\begin{align}
f(1, Y)=0,
\end{align}
where we used $\log 1=0$ and $\sin 0 =0$ for arbitrary value $Y$.
Note that when $Y\rightarrow 0$, then $\sqrt{\log X \log Y}$ is non-trivial.
However, due to $Y\rightarrow 0$, $f(1, Y)$ becomes $0$.
When $Y=1$,
\begin{align}
f(X, 1)=1-Y >0,
\end{align}
where we used $\log 1=0$ and $\sin 0 =0$ for arbitrary values $X$.
Note that when $X\rightarrow 0$, then $\sqrt{\log X \log Y}$ is also non-trivial.
However, in this case, $f(X, Y)$ is
\begin{align}
\lim_{X\rightarrow 0}f(X, Y)|_{Y=1}
=
1-\sin^2\left(\sqrt{\log X \log Y}\right)>0.
\end{align}
Thus, in case 1, $f(X, Y)$ is always positive. 
\\
\noindent
Case 2: $-X-Y\sin^2\left(\sqrt{\log X \log Y}\right)+Y=0$.
Then $f(X, Y)$ becomes
\begin{align}
f(X, Y)
&=
1-X-Y\sin^2\left(\sqrt{\log X \log Y}\right)
\nonumber\\
\quad
&=
1-X>0.
\end{align}
Thus, in case 2, $f(X, Y)$ is also always positive. 
In either case, $f(X, Y)\geq 0$, so the result is proven.
\end{appendix}

\end{document}